\documentclass[twocolumn,showpacs,preprintnumbers,amsmath,amssymb,prb,superscriptaddress]{revtex4-1}
\usepackage{graphicx}
\usepackage{dcolumn}
\usepackage{bm}

\begin{document}

\title{Influence of {\it e-e} scattering on the temperature dependence of the resistance of a classical ballistic point-contact in a 2DES}

\author{M.Yu.~Melnikov}
\affiliation{Institute of Solid State Physics, Russian Academy of
Sciences, 142432 Chernogolovka, Russian Federation}
\author{J.P.~Kotthaus}
\affiliation{Center for NanoScience and Fakult\"{a}t f\"{u}r Physik,
Ludwig-Maximilians-Universit$\ddot{\text{a}}$t,
Geschwister-Scholl-Platz 1, D-80539 M$\ddot{\text{u}}$nchen,
Germany}
\author{V. Pellegrini}
\affiliation{NEST, Istituto Nanoscienze-CNR and Scuola Normale Superiore, Piazza San Silvestro 12, I-56127 Pisa, Italy}
\author{L. Sorba}
\affiliation{NEST, Istituto Nanoscienze-CNR and Scuola Normale Superiore, Piazza San Silvestro 12, I-56127 Pisa, Italy}
\author{G.~Biasiol}
\affiliation{CNR-IOM, Laboratorio TASC, Area Science Park, I-34149 Trieste, Italy }
\author{V.S.~Khrapai}
\affiliation{Institute of Solid State Physics, Russian Academy of
Sciences, 142432 Chernogolovka, Russian Federation}

\begin{abstract}
We experimentally investigate the temperature ($T$) dependence of the resistance of a classical ballistic point contact (PC) in a two-dimensional electron system (2DES). The split-gate PC is realized in a  high-quality AlGaAs/GaAs heterostructure. The PC resistance is found to drop by more than 10\% as $T$ is raised from 0.5K to 4.2 K. In the absence of a magnetic field, the $T$-dependence is roughly linear below 2K, and tends to saturate at higher $T$. Perpendicular magnetic fields on the order of a few 10mT suppress the $T$-dependent contribution $\delta R$. This effect is more pronounced at lower temperatures, causing a crossover to a nearly parabolic $T$-dependence in magnetic field. The normalized magnetic field dependencies $\delta R(B)$ permit an empiric single parameter scaling in a wide range of PC gate voltages. These observations give strong evidence for the influence of electron-electron ({\it e-e}) scattering on the resistance of  ballistic PCs. Our results are in qualitative agreement with a recent theory of the {\it e-e} scattering based $T$-dependence of the conductance of classical ballistic PCs [Phys. Rev. Lett. {\bf 101}, 216807 (2008)] and [Phys. Rev. B {\bf 81}, 125316 (2010)].
\end{abstract}

\maketitle

\section{Introduction}

Electronic conductance of a diffusive solid-state system is characterized by a mean-free path of charge carriers ($l_0$). Scattering processes off disorder and phonons, as well as electron-electron scattering (via U-processes) contribute to a total quasi-momentum relaxation rate, inversely proportional to $l_0$. With increasing temperature ($T$) the scattering typically becomes more effective and the mean-free path decreases. The impact of scattering on the $T$-dependence of the conductance is different in ballistic systems, among which a classical point contact is, perhaps, the simplest.

Introduced by Sharvin~\cite{Sharvin}, the classical ballistic point contact (PC) is represented by an orifice between two clean reservoirs with dimensions large compared to the inverse Fermi-momentum ($k_F$). In two-dimensions (2D), the conductance of the classical PC is given by a well-known analogue of the Sharvin formula~\cite{Glazman}:
\begin{equation}\label{eq_Sharvin}
G_0= \frac{2e^2}{\hbar} \frac{ak_F}{\pi^2},
\end{equation}
where $e$ is the elementary charge, $\hbar$ is the Planck constant and $a$ is a half-width of the PC orifice, $ak_F\gg1$. Finite mean-free path in the reservoirs gives rise to a nonzero backscattering probability of the carriers injected through the PC. As a result, the PC conductance is smaller than the ideal value~(\ref{eq_Sharvin}) roughly by $\delta G/G_0\sim -a/l_0$. With increasing $T$ the mean-free path $l_0$ decreases and causes a negative $T$-dependence of the PC conductance. Important in disordered systems, this effect can be neglected in contemporary high quality devices ($a/l_0\sim1\%$) including those studied here.

In a recent work~\cite{NagaevAyvaz}, the impact of the {\it e-e} scattering on the conductance of classical PCs has been analyzed. As shown by the authors, the dominant contribution comes from scattering of the injected electrons with those incident onto the PC at large distances from the orifice. Contrary to naive expectations, this scattering mechanism {\it enhances} the conductance of the PC. For a classical PC in a two-dimensional electron system (2DES)
with Fermi energy $E_F$ and temperature $T$ the {\it e-e} scattering contribution is found to be (the numerical factor is given after~\cite{Nagaev_noise}):
\begin{equation}\label{eq_Ayvaz}
    \frac{\delta G_{ee}}{G_0}\approx 0.037\alpha_{ee}(ak_F)\frac{k_BT}{E_F}\ln(l_c/a),
\end{equation}
where $\alpha_{ee}\sim1$  is a dimensionless {\it e-e} interaction parameter, $k_B$ is the Boltzman constant and $l_c\sim l_0\gg a$ is a cutoff length-scale in the 2DES. For a typical PC in a 2DES in GaAs the {\it e-e} scattering contribution is estimated as $\sim10\%$ at liquid He temperatures and hence expected to dominate the $T$-dependence. More recently~\cite{NagaevKost}, a suppression of the {\it e-e} scattering contribution to the PC conductance in small perpendicular magnetic fields has been calculated.

Ballistic constrictions in 2D have been extensively studied since the first observations of conductance quantization~\cite{Wharam,Van_Wees}.
Thermal smearing of the conductance plateaus causes a well known $T$-dependence mechanism in such devices~\cite{Bagwell,Van WeesPRB1991}.
A quantum-dot-Kondo-like $T$-dependence in the region of the so-called 0.7 anomaly~\cite{point7} is associated with many-body physics.
E-e scattering-induced size effects in wide and long quasiballistic channels have been studied in~\cite{Molenkamp_ee}. In those experiments, a current self-heating was used to vary the electronic temperature and a non-monotonic behavior of the differential resistance was observed. The results were later interpreted as a 2D analogue of a hydrodynamic-like Gurzhi effect~\cite{gurzhi}. A non-local hydrodynamic-like problem of electron injection through a PC was addressed in~\cite{Govorov} in case of {\it e-e} scattering dominating all other scattering mechanisms. Corresponding experiments~\cite{schinner} have been performed in wide channels in a strongly nonlinear regime. Observation of the {\it e-e} scattering contribution to the linear response Sharvin conductance requires careful measurements of the $T$-dependence and magnetoresistance in PCs formed in a clean 2DES. We are aware of only one such experiment~\cite{renard}, which is qualitatively consistent with the {\it e-e} scattering scenario of Refs.~\cite{NagaevAyvaz,NagaevKost}.

Here we present an experimental study of the $T$-dependence of the resistance in split-gate defined PCs in a GaAs-based 2DES at liquid He temperatures.
In the absence of a magnetic field, a negative $T$-dependent resistance contribution on the order of 10\% is found. The $T$-dependence is suppressed by a small perpendicular magnetic field of a few 10mT. The normalized magnetic field dependencies for different gate-voltages obey an empiric single-parameter scaling procedure. The results give strong evidence for the influence of the {\it e-e} scattering on the PC conductance and are qualitatively consistent with the predictions of Refs.~\cite{NagaevAyvaz,NagaevKost}. We compare the measured functional dependencies on temperature and magnetic field with the theory predictions and discuss possible origins of the discrepancies, supporting by a numerical calculation. In the absence of a magnetic field a rough quantitative agreement is achieved and the interaction parameter $\alpha_{ee}$ is evaluated.

The paper is organized as follows. The experimental details are described in section~\ref{expdet}. The results in zero and finite
magnetic field are presented in sections~\ref{zerofield} and~\ref{Bfield}, respectively. Section~\ref{widthdep} is devoted to scaling of the magnetoresistance data. The discussion of the experimental results is given in section~\ref{disc}. The details of the evaluation of the electron density inside the PC and those of  numerical calculation are given in Appendix. The paper is briefly summarized in section~\ref{summary}.

\section{Experimental details\label{expdet}}

Our samples are based on a high quality (001)GaAs/AlGaAs heterostructure containing a 2DES 200nm below the surface. The electron density is
about $n_S\approx0.83\cdot10^{11} cm^{-2}$ (corresponding to $k_F\approx7.2\times10^5 cm^{-1}$) and the mobility is
$\approx4\cdot10^{6} cm^{2}/Vs$ at $T=4.2$K, which corresponds to an elastic mean free path of $\approx$20$\mu$m. The inset of fig.~\ref{fig1}a shows a micrograph of metallic split gates deposited on the crystal surface with the help of e-beam lithography (shown as two brighter areas). Negative gate voltages are applied to the gates in order to deplete the 2DES beneath them and define a PC. Throughout the paper, the voltage on the right gate is fixed at -0.4V, while the left gate voltage (below simply gate voltage, $V_g$) can be varied to control resistance of the PC. Using a finite bias spectroscopy~\cite{patel} we characterized a quasi-1D subband spacing ($\hbar\omega_\perp$) and a curvature of the PC potential along the transport direction ($\hbar\omega_\parallel$) in the spirit of a saddle-point model~\cite{buettiker}. This was done in a device similar to the one for which the experimental data are presented below. The subband spacing decreases appreciably with the number of occupied 1D channels: from $\hbar\omega_\perp\approx1.1$~meV for the 1-2 channels to $\approx$0.35~meV for the 5-6 channels. Respectively, the curvature along the transport direction decreases from $\hbar\omega_\parallel\approx0.55$~meV to $\approx$0.35~meV. These parameters reflect the experimental observation that the 5th and higher conductance plateaus are hardly resolved even at very low $T$~\cite{SM_QPC}. Consistently, the lithographic width of the PCs used is $260$nm, which gives an upper bound of 6 for the number of quantized conductance plateaus.

The ohmic contacts to the electron system are obtained via annealing of Ni/Au/Ge/Ni/Au layers and placed at
distances about 1mm away from the PC. The experiment is performed in a $^3$He insert at temperatures between 0.46 and 4.2 K and in (perpendicular to the interface) magnetic fields up to 330mT. The four-terminal resistance is measured with a lock-in amplifier with current excitation, ranging from 2 to 46nA at frequencies between 12 and 32Hz, strictly in the linear response. A 2DES resistance in series with the PC does not exceed 30$\Omega$ and has no impact on our results. The measurements were performed for a set of gate-voltages fixed during a cool-down.
The setup is equipped with a calibrated thermometer in the vicinity of the sample, which
allowed to measure the $T$-dependence of the PC resistance during a slow (more than 1 hour) cool-down or warm-up procedure.
Rare random jumps in the resistance on the order of 1\%, typically observed above 2K (see figs.~\ref{fig2}b and~\ref{fig5}a), are presumably caused by a nearby impurity recharging. When recording magnetic field dependencies, we compensated for these
by a minor shift ($<3\%$) of the curves in fig.~\ref{fig3}b, such that the zero field data points correspond to fig.~\ref{fig2}b.
Magnetic field sweeps exhibited a small hysteresis (less than $\pm10$mT), which was accounted  for with the help of Hall voltage measurements.
All together we measured three PCs on two samples prepared out of two similar wafers.
The results are practically sample independent and reproducible in respect to thermal recycling,
so that only the data for one PC obtained within the same thermal cycle are given below.
Throughout the paper, with a few exceptions, we present the data for the PC resistance, which was actually measured in experiment.
When comparing to theory, we make use of a relation between small contributions to the resistance and the conductance:
$\delta R/R_0=-\delta G/G_0$, with $R_0=1/G_0$.

\section{PC resistance in zero magnetic field\label{zerofield}}

\begin{figure}[t]
 \begin{center}
  \includegraphics[width=0.8\columnwidth]{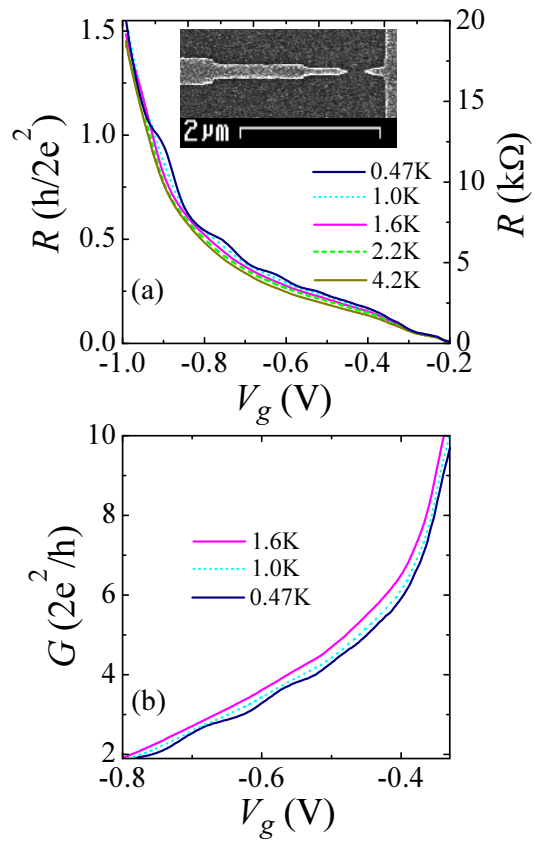}
   \end{center}
  \caption{Gate-voltage dependencies of the PC resistance/conductance. (a): a four-terminal PC resistance for a set of temperatures in absolute (right axis) and dimensionless (left axis) units. Inset: electron micrograph of the surface of a sample analogous to that used in the experiment. Two top gates arranged in a T-shaped geometry appear in a light-grey color. (b): a dimensionless PC conductance at the three lowest temperatures used in the experiment.}\label{fig1}
\end{figure}

Fig.\ref{fig1}a shows gate voltage dependencies of the PC resistance ($R$) for a set of
$T$ values. The traces start at $V_g=-0.2$V, where (and at more positive $V_g$) the device resistance $\approx30\Omega$ is dominated by that of the 2DES connected in series. At more negative gate voltage, the 2DES below the left gate depletes. Note, that the gate metallization consists out of three parts of different widths (inset of fig.~\ref{fig1}a), sometimes the width being smaller than the distance to the 2DES layer. That is why the 2DES depletion under the gate occurs within a relatively wide range $-0.22{\rm V}< V_g<-0.35$V, where three successive steps on the dependencies $R(V_g)$ can be resolved in fig.~\ref{fig1}a. Below $V_g\approx-0.35$V all the current flows through the PC constriction formed in the 2DES. The physical width of the PC, as well as the electron density inside it, is tuned by stray electric fields, such that its resistance  increases towards more negative $V_g$. Within the range $-0.5{\rm V}<V_g<-0.35$V the gate-voltage dependencies are featureless and the device behaves as a classical PC with the resistance $0.24 h/2e^2>R>0.12 h/2e^2$ (corresponding to $3.1 k\Omega>R>1.5 k\Omega$). In order to cover a broader resistance range, we also performed the experiment outside this gate-voltage interval. At more negative gate voltages $-0.8{\rm V}< V_g<-0.5$V we enter a quantum PC regime at the lowest $T$ and study the PC resistances up to $R\approx6.4k\Omega$ ($0.5 h/2e^2$). At more positive gate-voltages $-0.35{\rm V}< V_g< -0.27$V the PC constriction is not yet fully developed and its width exceeds the lithographic one. Here, PC resistances down to $R\approx0.64k\Omega$ ($0.05 h/2e^2$) are accessible. As seen already from fig.~\ref{fig1}a, the negative $T$-dependent contribution is found within the whole resistance range studied.

The effect of a finite temperature is easier to observe in the gate-voltage dependence of the PC conductance $G=1/R$. This is shown in fig.~\ref{fig1}b for a relevant $V_g$ range at the three lowest experimental temperatures. At the lowest $T\approx$0.47K the signatures of the quantized plateaus $2ne^2/h$ ($n=$2,3,4) are resolved. The impact of increasing $T$ is twofold: it causes (i) a smearing of the plateaus and (ii) an overall conductance increase.
The first effect is caused by a well-known thermal broadening~\cite{Bagwell} of the Fermi distribution function. However, as fig.~\ref{fig1}b demonstrates, at temperatures above $1K$ the gate-voltage dependencies of the conductance are already smooth, and this effect has a minor (if any) contribution to the $T$-dependence. In what follows, we focus on a second effect: the pronounced positive/negative $T$-dependence of the conductance/resistance, which is similar to the results of Ref.~\cite{renard} and observed within the whole gate-voltage interval in fig.~\ref{fig1}a and~\ref{fig1}b. This effect dominates the $T$-dependence above 1K.

\begin{figure}[t]
 \begin{center}
  \includegraphics[width=0.8\columnwidth]{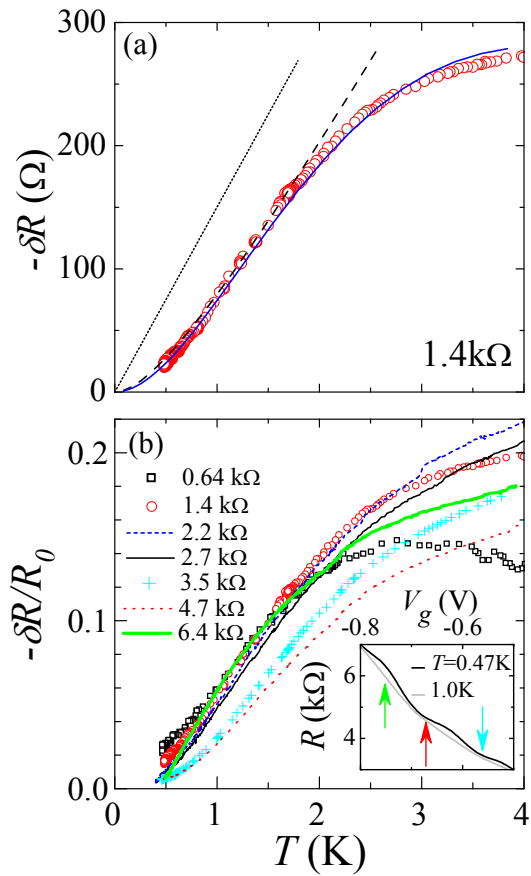}
   \end{center}
  \caption{Temperature dependent contribution to the PC resistance in the absence of a magnetic field. (a): experimental $T$-dependence for $R_0\approx1.4k\Omega$ in absolute units (symbols). Model fits are shown by lines, see section~\ref{disc_zeroB}. A fit according to eq.~(\ref{eq_Ayvaz}) for $\alpha_{ee}=0.77$, $a=2\times290$nm and $l_c=13\mu$m is shown by a dotted line. A result of a numeric calculation in the spirit of Ref.~\cite{NagaevAyvaz} is shown by a dashed line for the same parameters. This fit is intended to describe deviations from a linear $T$-dependence at low temperatures. The solid line demonstrates a numerical calculation which accounts for a beam decay effect owing to {\it e-e} scattering. The parameters are $\alpha_{ee}=1.75$, $l_c=5\mu$m and the same $a$. (b): normalized experimental $T$-dependencies  for a set of $R_0$ indicated in the legend. Inset: gate-voltage dependencies of the PC resistance for the two lowest temperatures. Arrows indicate the points where the three curves with high $R_0$ in (b) were taken.}\label{fig2}
\end{figure}

In fig.~\ref{fig2}a the $T$-dependence of the PC resistance is shown for $V_g=-0.337$V (symbols). The data are plotted in the form
of a $T$-dependent contribution $-\delta R\equiv -(R(T)-R_0)$, where $R(T)$ is the measured resistance at a given temperature. The resistance $R_0$, which  determines the absolute value of the $T$-dependent contribution is deduced from the magnetoresistance data of section~\ref{Bfield}. The overall $T$-dependence of the PC resistance in fig.~\ref{fig2}a is negative, such that $-\delta R$ monotonously increases with increasing temperature. The $T$-dependence is strongest
for $1K< T <2K$, where $-\delta R$ increases by a factor of 2.3 almost linearly
with $T$. At higher temperatures, the $T$-dependence becomes substantially weaker and $\delta R(T)$ tends to saturate.
Since $\delta R(T=0)=0$, the data imply that the $T$-dependence should also weaken at $T\rightarrow0$.
Yet the temperatures used are not low enough for this tendency to be pronounced in fig.~\ref{fig2}a.

Similar qualitative behavior is observed in a wide range of gate-voltages, as shown in fig.~\ref{fig2}b. Here the relative contribution $-\delta R/R_0$ is plotted as a function of $T$ for a set of PC resistances $0.64{\rm k\Omega}\leq R_0\leq6.4{\rm k\Omega}$. Again, the negative $T$-dependence is strongest at intermediate $T$, and weakens at high and low $T$, with some minor differences. For instance, the lowest resistance trace ($R_0=0.64k\Omega$) demonstrates a high-$T$ downturn, which seems to have the same origin as the saturation of the $T$-dependencies for higher $R_0$ (see section~\ref{disc_zeroB}). At $T<1$K  the three traces with high $R_0$ suffer from a residual thermal smearing of the $n=2,3,4$ resistance plateaus. This effect contributes a spurious positive (negative) $T$-dependence on the low-$V_g$ (high-$V_g$) edge of a resistance plateau~\cite{Bagwell}. As a result, below 1K the measured $T$-dependence is stronger for the $6.4k\Omega$ trace and weaker for the $4.7k\Omega$ and $3.5k\Omega$ traces, see the inset of fig.~\ref{fig2}b.

The data of fig.~\ref{fig2} demonstrate a negative $T$-dependence of the PC resistance on the order of $\sim10$\% at liquid He temperatures.
This effect is observed in a wide range of resistances from the limit of classical PC ($R_0\ll h/2e^2$) to the quantum PC regime (up to $R_0\approx 0.5h/2e^2$).
The negative $T$-dependence of the resistance cannot be interpreted in terms of a backscattering of electrons owing to a finite elastic mean-free path in the 2DES. In our ballistic PC, the overall backscattering contribution is too small (on the order of $a/l_0\sim1\%$) and would result in a positive $T$-dependence of the resistance. One can also safely neglect a possible contribution from the weak-localization in the 2D leads, which results in a negative $T$-dependent contribution several orders of magnitude smaller than that in fig.~\ref{fig2}. Instead of single-particle effects, we attribute the $T$-dependence of the PC resistance to the {\it e-e} scattering scenario~\cite{NagaevAyvaz}, which turns out to be very well consistent with the experiment (see section~\ref{disc}). Behavior of the magnetoresistance data in small perpendicular magnetic fields, presented in the next section, further supports this conjecture.

\section{PC resistance in a perpendicular magnetic field\label{Bfield}}

\begin{figure}[t]
 \begin{center}
  \includegraphics[width=0.8\columnwidth]{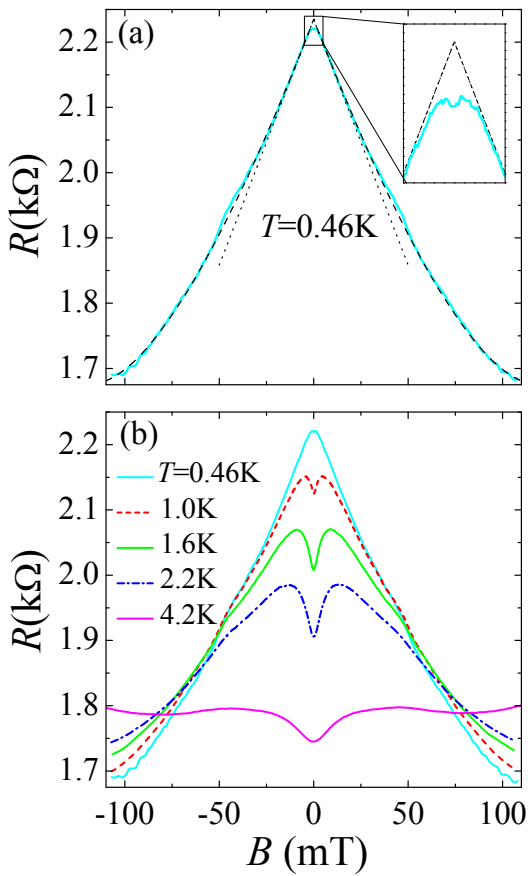}
   \end{center}
  \caption{Magnetoresistance of the PC for $R_0\approx2.2k\Omega$. (a): experimental dependence $R(B)$ for the lowest accessible $T$ (solid line) and the fits accounting for the suppression of backscattering in $B\neq0$ (dotted line) and, additionally, for the magnetoelectric subband depopulation (dashed line). Inset: magnified low-$B$ region of the main figure with dimensions 45$\Omega\times10$mT. The experimental data in the inset (solid line) are obtained in a separate low-$B$ sweep with a better resolution. The fits in (a) are obtained assuming the electron densities of $8.3\cdot10^{10}cm^{-2}$ and $3.15\cdot10^{10}cm^{-2}$, respectively, in the 2DES and inside the PC. The deduced $R_0$ equals 2235$\Omega$.  (b): evolution of the experimental magnetoresistance with temperature.}\label{fig3}
\end{figure}

In fig.~\ref{fig3}b the experimental PC magnetoresistance is plotted for a set of temperatures at a gate-voltage of $V_g=-0.41$V, corresponding to a classical PC regime. In small fields, the curves $R(B)$ demonstrate a negative $T$-dependence. The dependence is strongest at $B=0$ and quickly suppressed with increasing magnetic field. As a result, a pronounced zero-field resistance minimum is observed, which becomes wider and deeper as $T$ is raised. The $T$-dependent magnetoresistance is well seen even for $|B|<$10mT, where the cyclotron radius ($R_C>4\mu$m) by far exceeds the size of the PC orifice. This observation gives a strong evidence that the $T$-dependent PC magnetoresistance originates from the neighboring 2DES in agreement with the {\it e-e} scattering scenario~\cite{NagaevKost}. In higher magnetic fields $|B|\gtrsim50$mT a sign change of the $T$-dependence is observed in fig.~\ref{fig3}b (signatures of such a behavior were also seen in~\cite{renard}), which we failed to unambiguously interpret. For this reason, in the following we limit our analysis to the magnetic field range $|B|\leq25$mT.

Before analyzing the $T$-dependent contribution in detail, we note, that in fig.~\ref{fig3}b it is superimposed on a well-known single-particle magnetoresistance~\cite{vanHouten_beenakker} (see also~\cite{NagaevKost}). The latter contribution is independent of $T$, and can be identified  in the low-$T$ limit. In fig.~\ref{fig3}a the PC resistance at the lowest $T\approx0.47$K is plotted as a function of $B$ (thick line) for the same $V_g$ as in fig.~\ref{fig3}b. The overall magnetoresistance is negative and well explained by two single-particle effects. The first contribution results from a non-additivity of the ballistic PC resistance and Hall resistance $R_H=B/(ecn_S)$ of the neighboring 2DES leads (so-called suppression of backscattering~\cite{suppression_backscattering}). In a quasiclassical approximation~\cite{vanHouten_beenakker}, the four-terminal PC resistance exhibits a negative linear magnetoresistace $\delta R(B)=-|R_H|\propto-|B|$. As seen from fig.~\ref{fig3}a, for $|B|\leq25$mT the experimental magnetoresistance has the same slope as that of $-|R_H|$ (dots) with a value of $n_S$ deduced from Shubnikov oscillations in the bulk 2DES. In stronger magnetic fields, the experimental magnetoresistance weakens, owing to a second effect --- namely, depopulation of magnetoelectric subbands in the PC~\cite{vanHouten_beenakker}. To the lowest order in $B$, this effect contributes a positive magnetoresistance $\propto B^2$. As shown in fig.~\ref{fig3}a, taking both these contributions into account results in a nearly perfect fit (dashed line) to the experimental curve for $|B|\leq100$mT. Hence, other possible $B$-dependent contributions (including {\it e-e} scattering~\cite{NagaevKost}) are negligible at this temperature, except for a tiny field range magnified in the inset of fig.~\ref{fig3}a.

In the inset of fig.~\ref{fig3}a, the experimental magnetoresistance is shown along with the fits of fig.~\ref{fig3}a extrapolated towards $|B|\rightarrow0$. At $B=0$, the single-particle extrapolation predicts a resistance slightly higher than actually measured. In part, this deviation is caused by a rounding of the sharp $B=0$ kink in a real 2DES~\cite{roundingfig3a}. However, a tiny region of very weak positive magnetoresistance seen in the inset of fig.~\ref{fig3}a
suggests that at the lowest temperature used the $T$-dependent contribution survives for $|B|<3$mT and contributes to the deviation from the extrapolated single-particle fits. This deviation being small ensures that the complete fit to the lowest $T$ magnetoresistance data (dashed line in fig.~\ref{fig3}) can serve as a reliable $T\rightarrow0$ limit~\cite{WL}. This fit is used as a reference curve $R^{REF}(B)$ to determine the $T$-dependent contribution $-\delta R(B,T)=R^{REF}(B)-R(B,T)$, where $R(B,T)$ is the measured resistance at a given magnetic field and temperature. In zero field we have $R_0\equiv R^{REF}(B=0)$. The same procedure is performed at every gate-voltage value. All such fits are performed with the same Hall coefficient, corresponding to the bulk 2DES density. The subband depopulation effect in each case is fitted separately in a hard-wall model with an electron density inside the PC as a fit parameter (Appendix~\ref{depopulation}). A possible systematic error ($<1$\% in the classical PC regime) in chosen $R^{REF}(B)$ is not important as long as $|\delta R|/R_0$ is not too small, i.e. for $T\geq1$K and in not too high magnetic fields. Note, that in the quantum PC regime such a procedure does not allow to overcome the $T$-dependence owing to the smearing of the resistance plateaus, which is independent of magnetic field. Hence the systematic error is higher in that case and on the order of a few \%.

\begin{figure}[t]
 \begin{center}
  \includegraphics[width=0.8\columnwidth]{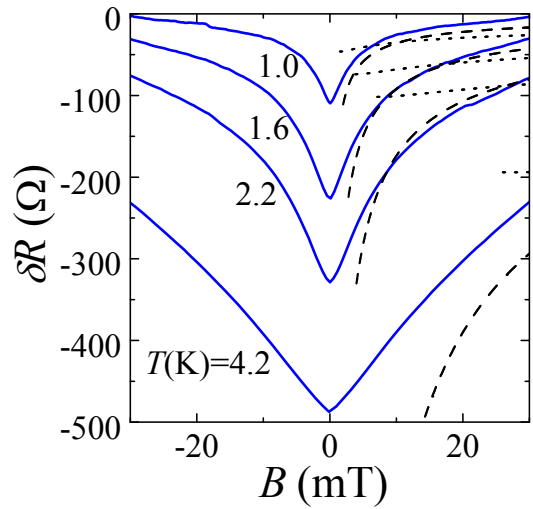}
   \end{center}
  \caption{Magnetic field suppression of the $T$-dependent correction to the PC-resistance. Experimental data (obtained from fig.~\ref{fig3}b, see text) are shown with solid lines. The best fit with eq.~(\ref{eq_kost}) (parameters $\alpha_{ee}=0.3$, $a=204$nm, $R_0=2235\Omega$ with an assumed density inside the PC $3.15\cdot10^{10}cm^{-2}$) and the empiric fit $\delta R_{ee}\propto T^2|B|^{-0.7}$ are shown with a dotted and dashed line, respectively.}\label{fig4}
\end{figure}

In fig.~\ref{fig4} the $B$-dependencies of $\delta R(B,T)$, obtained from the data of fig.~\ref{fig3}b as described above, are plotted for a set of temperatures. The absolute value $|\delta R|$ exhibits a maximum at $B=0$, symmetric with respect to field reversal and decays in magnetic field. The functional $B$-dependence weakens as the magnetic field is increased. Dashed lines in fig.~\ref{fig4} show the best empiric power law fit to the experimental data set with an expression $\delta R\propto -T^2B^{-0.7}$. We find such a fit reasonable at intermediate temperatures $T\leq 2.2$K except for small $|B|$. Scaling of the magnetoresistance data at different $V_g$ (section~\ref{widthdep}) indicates that similar functional $B$-dependencies $-\delta R(B)$  are observed in a wide range of the PC gate voltages.

\begin{figure}[t]
 \begin{center}
  \includegraphics[width=0.8\columnwidth]{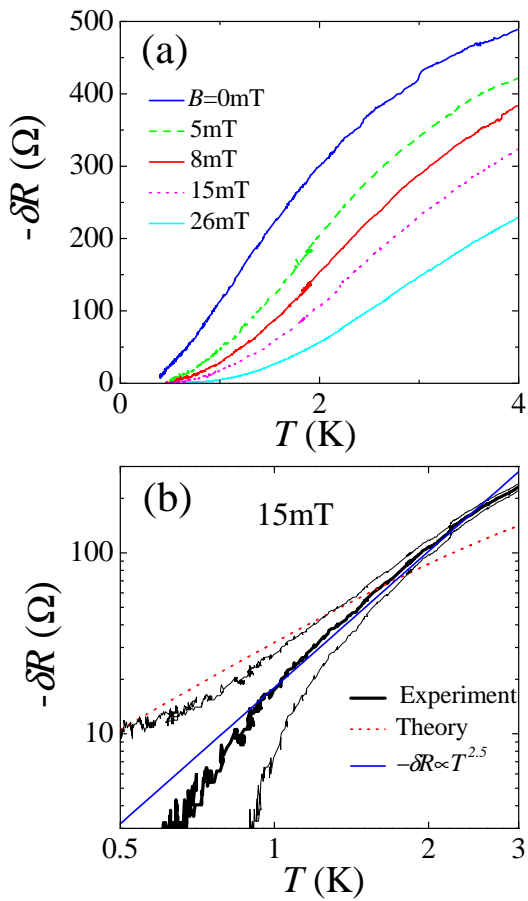}
   \end{center}
  \caption{Evolution of the $T$-dependence with magnetic field for $R_0\approx2.2k\Omega$. (a): experimental $T$-dependencies plotted in linear scale for a set of magnetic field values. (b): double-log plot of the experimental dependence at $B=15$mT (thick solid line), together with the fit to eq.~(\ref{eq_kost}) (parameters $\alpha_{ee}=0.3$, $a=204$nm, $R_0=2235\Omega$ with an assumed density inside the PC $3.15\cdot10^{10}cm^{-2}$) and the empiric dependence $\delta R\propto-T^{2.5}$. Thin solid lines indicate the range of possible systematic uncertainty associated with the choice of $R_0$. These lines are obtained via shifting the thick solid line by $\pm10\Omega$. }\label{fig5}
\end{figure}

In fig.~\ref{fig5}a the measured $T$-dependencies are plotted for a set of fixed (positive) $B$ values. At $B=0$, as discussed above, a nearly linear $T$-dependence is observed for $T<2$K, followed by saturation at higher $T$. The $B$-driven suppression of the $T$-dependent contribution is stronger at lower $T$. As a result, with increasing $B$ the $T$-dependence approaches parabolic at low temperatures.

For a quantitative analysis of the $T$-dependence we plot the data at $B=15$~mT on a double logarithmic scale in fig.~\ref{fig5}b (thick solid line). The experimental uncertainty  is indicated as follows. The reference value $R^{REF}(B)$ is determined with an accuracy of $\pm10\Omega$, appreciably worse than the random error in a measured $T$-dependence ($\pm2\Omega$). This unknown additive contribution to $\delta R(T)$ could shift the data up or down as a whole, to a position somewhere between those indicated with thin solid lines in fig.~\ref{fig5}b. Within this uncertainty, the experimental data is compatible with a power-law $\delta R(T)\propto-T^\alpha$, where $\alpha\approx2.5\pm0.5$. The best such fit
$-\delta R\propto T^{2.5}$ is shown by a thin solid line in fig.~\ref{fig5}b.

The key experimental observations of figs.~\ref{fig3},~\ref{fig4} and~\ref{fig5} can be qualitatively explained in the framework of the {\it e-e} scattering contribution to the PC resistance. As predicted in Ref.~\cite{NagaevKost}, the {\it e-e} scattering is suppressed by a perpendicular magnetic field $B\neq0$, as a result of time-reversal symmetry breaking. A typical $B$-field scale (a few 10~mT) is defined by competition of an interaction time argument~\cite{NagaevKost} and phase-space constraints and increases with $T$. The corresponding magnetoresistance contribution is positive and $T$-dependent. It is predicted to be observable already in tiny magnetic fields, where the cyclotron radius in the 2DES $R_C=\hbar k_Fc/eB$ is much larger than the size of the orifice~\cite{NagaevKost}, in perfect agreement with our experiment. Next we demonstrate that the magnetoresistance data are similar in a wide range of the PC gate-voltages (i.e. $R_0$), permitting a single parameter scaling.

\section{Scaling the magnetoresistance data at different gate-voltages\label{widthdep}}

\begin{figure}[t]
 \begin{center}
  \includegraphics[width=0.8\columnwidth]{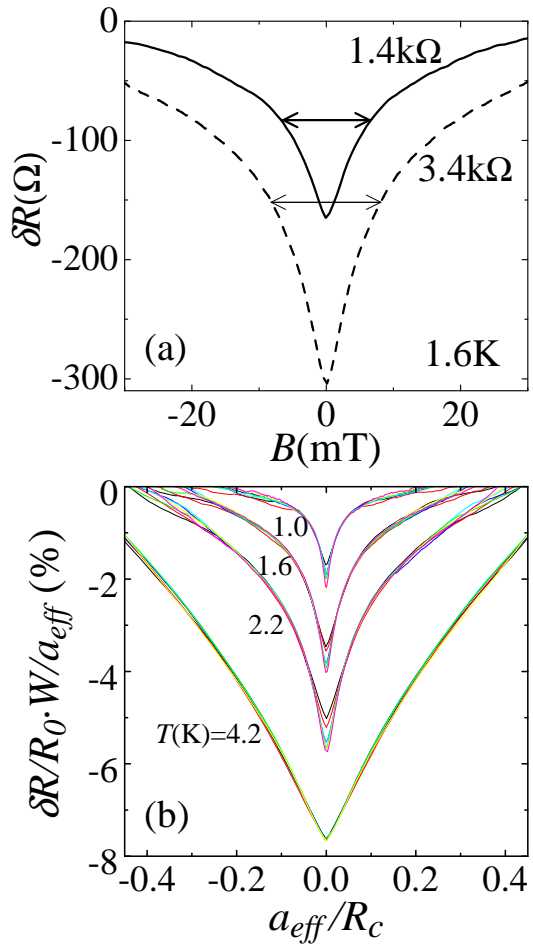}
   \end{center}
  \caption{Magnetoresistance at different $R_0$. (a): experimental dependencies $\delta R(B)$ for two values of $R_0$ at 1.6K. The arrows mark the full width at the half minimum of the V-shaped dependencies. (b): One-parameter scaling of the magnetoresistance data for $R_0\leq6.4k\Omega$. The temperature increases from top to bottom $T=$1, 1.6, 2.2, 4.2K.}\label{fig6}
\end{figure}

In fig.\ref{fig6}a we compare the curves $\delta R(B)$ taken at $T=1.6$K for two different $V_g$ corresponding to the PC resistances of $R_0\approx1.4k\Omega$ and  $R_0\approx3.4k\Omega$. In both cases, the overall effect of magnetic field is qualitatively similar to that in fig.~\ref{fig4} and the absolute value $|\delta R(B)|$ increases with $R_0$. At the same time, the width of the V-shaped dependence along the $B$-axis
also increases with $R_0$, as demonstrated by arrows in fig.~\ref{fig6}a, which mark the full width at the half minimum $\delta R(B)$. This behavior is observed within the whole range of $R_0$ investigated and for all $T$ used.

We find that the data $\delta R(B)$ at different $R_0$ permit a single parameter scaling. At a given $T$, an empiric relation
$a_{eff}^{-1}\delta R/R_0\approx F(a_{eff}B)$ holds, where $a_{eff}$ is an $R_0$-dependent scaling parameter and $F$ is a function which determines the shape of the $B$-dependence at this $T$. It is convenient to express $a_{eff}$ in units of length and plot the scaled $B$-dependencies in dimensionless coordinates $\delta R/R_0\cdot w/a_{eff}$ vs $\beta\equiv a_{eff}/R_C$, where $w=130$nm is the lithographical half-width of the PC and $R_C\propto B^{-1}$ is the cyclotron radius in the 2DES. This choice of axes is natural for the {\it e-e} scattering mechanism of the $T$-dependence, as explained in section~\ref{disc}.

In fig.~\ref{fig6}b, we plot the results of the scaling procedure for a wide range of $R_0$. For each of the four temperatures used, the data are obtained as follows. We vary the value of $a_{eff}$ such that the dependencies $\delta R(B)/R_0$ at different $R_0$ fall on a single curve. For better scaling, the value of $R_0$ is also slightly varied within the experimental uncertainty (a few \%). Note that the values of $R_0$ are consistent with the $T=0.47$K data~\cite{SM_QPC} and are not expected to exhibit clear resistance quantization effects for the reasons discussed above (see section~\ref{Bfield}). At a given $T$, a nearly perfect scaling is obtained in the region of not too high $|\beta|$, where $\delta R(B)/R_0$ is not small, and outside the very vicinity of $\beta=0$ (i.e. zero magnetic field). The scaling is successful even far from the classical PC limit, up to $R_0\approx 0.5h/2e^2$. Remarkably, a similar low-$B$ behavior is observed even close to the resistance quantum $h/e^2$, although the scaling is worse for such high $R_0$ (not shown).

\begin{figure}[t]
 \begin{center}
  \includegraphics[width=0.8\columnwidth]{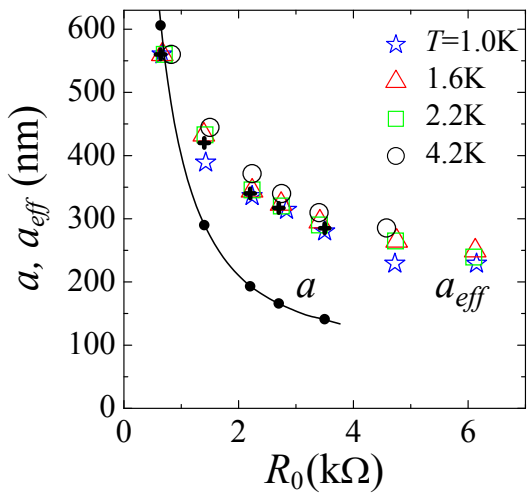}
   \end{center}
  \caption{Dependence of the scaling parameter $a_{eff}$ from fig.~\ref{fig6} (symbols) and the physical half-width $a$ of the PC (line) on $R_0$. The physical half-width $a$ is evaluated from eq.~(\ref{eq_Sharvin}) with $G_0=1/R_0$, with a reduced electron density inside the PC taken into account (see Appendix~\ref{depopulation}). The scaling parameter $a_{eff}$, normalized as described in the text, is shown by separate symbols for the four temperatures used. Black dots and crosses mark, respectively, the values of $a$ and $a_{eff}$ used in numerical fits, see Appendix~\ref{numerics}.}\label{fig7}
\end{figure}

In fig.~\ref{fig7} we plot the scaling parameter $a_{eff}$ (symbols) as a function of $R_0$. The absolute value of $a_{eff}$ is not determined in a scaling procedure and is chosen to coincide with the physical half-width $a$ of the PC on the lowest $R_0$ end. Normalized in this way, $a_{eff}$ is nearly $T$-independent and decreases at increasing $R_0$. As seen from fig.~\ref{fig7}, $a_{eff}$ drops by less than a factor of 3 when $R_0$ is varied by almost an order of magnitude.

\section{Discussion\label{disc}}

As demonstrated in previous sections, the resistance of a ballistic PC in a high-quality 2DES exhibits a negative $T$-dependence at liquid He temperatures. In the absence of magnetic field, the $T$-dependence is relatively strong below $T\approx2$K and tends to saturate at higher temperatures.
At not too high $T$, a small magnetic field on the order of a few 10mT tends to suppress the $T$-dependence. These observations can be well explained
in terms of the {\it e-e} scattering contribution to the ballistic PC resistance. The experimental data are qualitatively consistent with the predictions of the theory of Refs.~\cite{NagaevAyvaz,NagaevKost} for classical PCs, which is the only present theory capable to closely describe the data. Below we discuss the experimental results in detail, quantitatively compare the experimental behavior with key theoretical predictions and attempt to evaluate the strength of the {\it e-e} scattering in the 2DES studied.

\subsection{Zero magnetic field\label{disc_zeroB}}

\paragraph{Qualitative picture at not too high $T$.}
At temperatures in the range $0.5{\rm K}<T\lesssim2$K, the experimental $T$-dependence of the resistance is not far from linear (see symbols fig.~\ref{fig2}a). Still, at low $T$ the dependence is incompatible with the proportionality $|\delta G_{ee}|\propto T$ predicted by eq.~(\ref{eq_Ayvaz}).
This is strongly supported by the magnetoresistance data of section~\ref{Bfield}, where the contribution of {\it e-e} scattering is almost negligible at the lowest temperature used $T\approx0.5$K. This behavior is captured  by a numerical calculation of the {\it e-e} scattering contribution in spirit of Ref.~\cite{NagaevAyvaz} (see Appendix~\ref{numerics} for the details) and can be qualitatively understood as follows.

Consider a scattering of an injected electron (momentum ${\bf k}$) with the one incident onto the PC (momentum ${\bf p}$) at large distances $r$ from the orifice ($r\gg a$). The two momenta are approximately opposite, such that the angle $\varphi$ between ${\bf k}$ and ${-\bf p}$ is small: $|\varphi|\leq\phi_{PC}\ll1$, where $\phi_{PC}=a/r$ is the angular dimension of the PC orifice at a distance $r$. For $|\varphi|\leq\phi_T$ (with a thermal angle defined as $\phi_T\approx T/E_F$), the two electrons can scatter by an arbitrary angle~\cite{gurzhi} and the scattering probability is independent of $\varphi$. Outside this range, for $|\varphi|>\phi_T$, the scattering probability decreases with increasing $|\varphi|$ and, accordingly, the scattering angle cannot exceed $\sim\phi_T/|\varphi|$. These constraints result from the conservation laws and the Pauli principle. Depending on $T$ and/or $r$, two limiting cases can be realized.

\begin{figure}[t]
 \begin{center}
  \includegraphics[width=0.6\columnwidth]{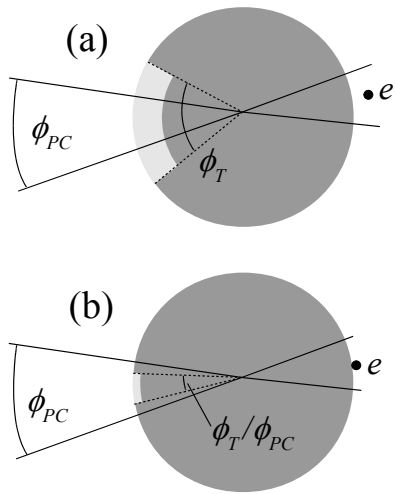}
   \end{center}
  \caption{A sketch of momentum space in the 2DES at a distance $r\gg a$ from the orifice for two limiting cases of not too low $T$ (or large $r$: $\phi_T\gg\phi_{PC}$) (a) and very low $T$ (or small $r$: $\phi_T\ll(\phi_{PC})^2$) (b), see text. A typical state of a non-equilibrium electron injected through the PC is shown by a black dot. Dark grey color indicates occupied states below the Fermi surface. Electron states with approximately opposite momenta available for {\it e-e} scattering are shown as light grey areas (see text for the details).}\label{ee_sketch}
\end{figure}

Fig.~\ref{ee_sketch} sketches the momentum space for the two limiting cases. A situation of fig.~\ref{ee_sketch}a is realized when $\phi_{PC}<\phi_T$,
i.e. at not too low $T$ (and/or not too small $r$). In this case, an average injected electron (shown by a black dot) can scatter by an arbitrary angle with each of the electrons within the stripe $|\varphi|<\phi_T$ of the width $\sim T$ (shown by a light grey color). However, only the scattering with the electrons within the angle $|\varphi|\lesssim\phi_{PC}$ can contribute to the PC conductance. Hence, $\delta G_{ee}\sim \phi_{PC}\times T$, which in the end gives rise to the linear $T$-dependence and a log-dependence on $l_c$ in eq.~(\ref{eq_Ayvaz}), derived in Ref.~\cite{NagaevAyvaz}. At $T\rightarrow0$ the condition $\phi_{PC}<\phi_T$ breaks down and one expects deviations from the linear $T$-dependence. In fig.~\ref{ee_sketch}b we sketch a situation in  the limit of very low $T$, such that  $\phi_T\ll(\phi_{PC})^2$. Here an average injected electron (black dot) cannot effectively scatter an incident electron unless $|\varphi|\lesssim\phi_T/\phi_{PC}$. The corresponding momentum space region of width $\propto T$ is shown by a light grey color in fig.~\ref{ee_sketch}b. Outside this region, for larger $|\varphi|$, the typical scattering angle is smaller than $\phi_{PC}$. Corresponding scattering processes do not influence the conductance for one of the scattered electrons still reaches the PC orifice. In this limit the {\it e-e} scattering contribution acquires a parabolic~\cite{super_finding} $T$-dependence $\delta G_{ee}\propto T^2$.

The dashed line in fig.~\ref{fig2}a represents a numerical fit to the experimental data (symbols) performed in the framework of Ref.~\cite{NagaevAyvaz}.
The calculation exhibits a crossover between the two limiting cases considered above and is quantitatively consistent with the experiment at $T\leq2$K for the chosen fit parameters (see caption). For comparison, the dependence predicted by eq.~(\ref{eq_Ayvaz}) for the same fit parameters is shown by the dotted line in fig.~\ref{fig2}a. Note that the numerically calculated $T$-dependence is strongly nonlinear below 0.5K. This demonstrates that a naive linear extrapolation of the experimental dependence $R(T)$ towards $T=0$ may be misleading and an independent determination of $R_0$ is needed (see section~\ref{Bfield}). The low-$T$ behavior is related to the value of the cutoff length $l_c$, which is expected to be comparable to the elastic mean-free path~\cite{NagaevAyvaz}. As discussed in Appendix~\ref{numerics}, at increasing $l_c$ the numerical result asymptotically approaches the analytic prediction of eq.~(\ref{eq_Ayvaz}) for arbitrarily low temperatures. Hence, a proportionality~\cite{NagaevAyvaz} $\delta G^{ee}\propto T$ might be observed in samples of exceptionally high quality in future (see Appendix~\ref{numerics}).

\paragraph{High-$T$ behavior and beam decay effects.}
As seen from fig.~\ref{fig2}a, the experimental $T$-dependence  levels off and tends to saturate at $T>2$K. A similar  behavior $\delta R(T)$ is observed within the whole range of gate-voltages studied (fig.~\ref{fig2}b), which is not predicted by theory~\cite{NagaevAyvaz}.
We find that such a qualitative behavior is in fact expected for the {\it e-e} scattering scenario at high enough $T$ and is related to a breakdown of a condition $l_c\ll l_{ee}$, where $l_{ee}$ is an electron mean-free path for {\it e-e} scattering. In such a situation, a decay of the injected beam on the length scale of $l_{ee}$ owing to the {\it e-e} scattering is not negligible. Therefore, the total {\it e-e} scattering probability is no longer small and the calculations~\cite{NagaevAyvaz} to first order in $\alpha_{ee}$ are expected to overestimate the effect of {\it e-e} scattering on the PC conductance.

One can estimate the length $l_{ee}$ based on a calculation of a quasiparticle lifetime~\cite{giuliani}. This calculation agrees with energy-relaxation data in a high-density device~\cite{leroy}. In our 2DES at $T=4$K such an estimate gives $l_{ee}\approx2.7\mu m$, indeed, much smaller than the elastic mean-free path $l_0$. Moreover, we estimate that the condition $l_0\sim l_c\ll l_{ee}$ is satisfied only at $T\ll2$K. Therefore, the applicability range of the numeric calculation in spirit of Ref.~\cite{NagaevAyvaz} turns out too narrow for a reliable estimation of the {\it e-e} interaction parameter $\alpha_{ee}$. Unfortunately, a mathematically strict account for the effects of the beam decay owing to {\it e-e} scattering
is too involved for modeling. Nevertheless, we performed additional numerical calculations with an ad-hoc introduced decay of the injected beam with characteristic length-scale of $l_{ee}$. These calculations are capable of reproducing the $T$-dependence in the whole temperature range studied, see Appendix~\ref{numerics}. One such fit perfectly consistent with the experiment is plotted by a solid line in fig.~\ref{fig2}a. Note that the interaction parameter $\alpha_{ee}$ obtained from such fits is roughly 2 times larger compared to the fit in which the beam decay effect is disregarded (dashed line in fig.~\ref{fig2}a).

\subsection{Finite magnetic field\label{disc_Bfield}}
Thanks to the time-reversal symmetry breaking, the contribution of the {\it e-e} scattering to the PC conductance is expected to rapidly decay in a perpendicular magnetic field~\cite{NagaevAyvaz}. The theoretical prediction of Ref.~\cite{NagaevKost} for not too small magnetic fields is given by:
\begin{equation}\label{eq_kost}
    \frac{\delta G_{ee}}{G_0}=\frac{2}{9}\alpha_{ee}\frac{ak_F}{\sqrt{\beta}}\frac{k_B^2T^2}{E_F^2}\ln\left(\beta\frac{E_F^2}{k_B^2T^2}\right).
\end{equation}
This result is derived assuming $k_B^2T^2/E_F^2\ll\beta\ll1$, where $\beta\equiv a/R_C\propto B$ is the ratio of the half-width of the PC orifice to the cyclotron radius in the 2DES $R_C=\hbar k_Fc/eB$.

Eq.~(\ref{eq_kost}) predicts that a decay of $\delta G_{ee}$ in a perpendicular magnetic field causes a $T$-dependence stronger than that at $B=0$.
On the other hand, as follows from this expression, at higher temperatures a decay of $\delta G_{ee}$ occurs in stronger magnetic fields scaling as $\sqrt{\beta}\propto T$. This behavior has to do with the physics of {\it e-e} scattering in perpendicular magnetic field. Quasiclassically, the {\it e-e} scattering probability is proportional to the effective interaction time between the scattering electrons. The interaction time depends on the angle
$\varphi$ between the momenta ${\bf k}$ and ${-\bf p}$ (defined in section~\ref{disc_zeroB}). In $B=0$ case, this time is maximum for electrons with opposite momenta ($\varphi=0$), which follow time-reversed trajectories. As derived in Ref.~\cite{NagaevKost}, in finite $B$ the time-reversal symmetry is broken and the maximum interaction time is attained for scattering of electrons with a nonzero angle: $|\varphi|\propto\sqrt{\beta}$. Along with the phase space argument of section~\ref{disc_zeroB} this explains the $B-T$ scaling mentioned. The low-$B$ experimental data of figs.~\ref{fig4} and~\ref{fig5}a are in perfect qualitative agreement with these theoretical predictions, strongly supporting our interpretation in terms of the {\it e-e} scattering scenario of Refs.~\cite{NagaevAyvaz,NagaevKost}.

The functional $T$- and $B$- dependencies predicted by expression~(\ref{eq_kost}) are weaker than observed in experiment.
Dotted lines in fig.~\ref{fig4} represent an effort to fit the experimental $B$-dependence with expression~(\ref{eq_kost}) in the range of magnetic fields where it's applicable. Despite the overall trends are qualitatively captured by the fits, we observe that a functional $B$-dependence in the experiment is appreciably stronger than the theory predicts. Except for the highest $T=4.2$K, the empiric dependence of the form $\delta R\propto T^2B^{-0.7}$ describes the data much better in the same range of $B$ (dashed lines in fig.~\ref{fig4}). The $T$-dependence in magnetic field is also somewhat stronger than the theoretical prediction. The dashed line in fig.~\ref{fig5}b, drawn according to eq.~(\ref{eq_kost}), is not consistent with the experimental dependence (thick solid line) in a typical magnetic field of $B=15$mT. At the same time, the empiric dependence $|\delta R(T)|\propto T^{2.5}$ can well describe the data within the experimental uncertainty.  Note, that the $T$-dependence of fig.~\ref{fig5}b is discussed in a limited range $T\leq3K$, which is related to a possible problem with the applicability of the theoretical analysis at high temperatures (see section~\ref{disc_zeroB}).

Several possibilities might be responsible for the discrepancy between the theory and experiment in magnetic field. First, we mention the unknown positive $T$-dependence of the resistance seen for $|B|\gtrsim50$mT (fig.~\ref{fig3}). If additive with the {\it e-e} scattering induced magnetoresistance at lower $|B|$, such a "wrong-sign" contribution can strengthen the dependence $\delta R(B)$. Second, the T-shaped split-gate layout of our sample (inset to fig.~\ref{fig1}) is different from the theoretical geometry~\cite{NagaevKost}, which might be important for the calculation of the effective interaction times used in the theory. Finally, long-range density gradients around the lateral PC and beam collimation effects discussed in the next subsection can well affect the {\it e-e} scattering contribution in $|B|\neq0$.

\subsection{Scaling and $a_{eff}$}

As demonstrated in section~\ref{widthdep}, the experimental dependencies $\delta R(B)$ at different $R_0$ obey scaling in the form $a_{eff}^{-1}\delta R(B)/R_0=F(a_{eff}/R_C)$, where the function $F$ determines the shape of the $B$-dependence at a given temperature. Here we argue that such a scaling is inherent for the {\it e-e} scattering problem in a 2DES in perpendicular magnetic field.

Consider two classical PCs of widths $a_1$ and $a_2$ in magnetic fields corresponding to cyclotron radii of, respectively, $R_1$ and $R_2$. Since the phase space for {\it e-e} scattering is determined essentially by the angle between the momenta of scattering electrons, these two problems are similar and can be reduced to each other via scaling the spatial coordinates, provided $a_1/R_1=a_2/R_2$. The only difference is that the total scattering probability (proportional to the interaction time) scales as $a_i$. Hence, we conclude that the {\it e-e} scattering contribution obeys a scaling in the form $a^{-1}\delta G^{ee}/G_0=F(a/R_C)$. It appears that both analytical expressions of Ref.~\cite{NagaevKost} derived in the limits of low and high $\beta$ obey such a scaling. This universality is expected to break down only at very low magnetic fields ($R_C\gg l_c$), where a weak disorder scattering comes into play~\cite{NagaevKost}.

As shown in fig.~\ref{fig6}b, the experimental dependencies $\delta R(B)$ permit scaling of this kind in a wide range of $R_0$ and all $T$ used. This further supports our interpretation of the negative $T$-dependence in small $B$ in terms of the {\it e-e} scattering mechanism~\cite{NagaevKost}. The scaling also indicates that a problem with a functional dependence $\delta R(B)$, which is appreciably stronger in experiment than in theory, persists regardless $R_0$.

As follows from the above argument, the scaling parameter is proportional to the PC half-width $a$. Unexpectedly, the evolution of $a_{eff}$ with $R_0$ in fig.~\ref{fig7} is appreciably weaker than that of the physical half-width $a$ (symbols and line, respectively). This discrepancy could not be ascribed to experimental uncertainties in scaling procedure and/or evaluation of $a$ (see Appendix~\ref{depopulation}). The data of fig.~\ref{fig6}b and fig.~\ref{fig7} suggest that a different length-scale $\propto a_{eff}$, rather than $a$, controls the behavior of the {\it e-e} contribution to the resistance of a laterally defined PC in a magnetic field. This behavior might be related to a long-range potential landscape nearby the PC, as briefly discussed below.

As a result of stray electric fields from the split-gates, an electrostatically defined PC takes the shape of a finite length channel with smooth boundaries. This causes a classical beam collimation effect in lateral PCs~\cite{Molenkamp}. A reduced electron density inside the PC also results in beam collimation via an electrostatic lens effect~\cite{vanHouten_beenakker,sivan,spector}. In a model of a horn-shaped channel~\cite{vanHouten_beenakker}, the angular dimension of the collimated (injected or incident) electron beam is reduced compared to $\pi$. This is compensated by the increased channel width $a_{exit}$ at the exit (entrance) of the PC, although the conductance $G_0$ is still given by the minimum channel width $a$ and eq.~(\ref{eq_Sharvin}). Hence, the angular dimension of the PC observed at large distances is increased: $\phi_{PC}\propto a_{exit}>a$, which is expected to enhance the relative {\it e-e} scattering contribution. This effect resembles the operation of a microwave horn antenna. In addition, the collimated beam is squeezed towards the normal to orifice, so that the angular dimension of the orifice is, on the average, further increased: $\overline{\phi_{PC}}\propto\overline{\cos\phi_n}$, where $\phi_n$ is the angle in respect to the normal to the orifice. Obviously, both effects become more pronounced as the PC is being depleted. One can speculate, that at large $l_c$ these effects might be roughly described by an extra effective width parameter $a_{eff}>a$ used instead of $a$ in eq.~(\ref{eq_Ayvaz}). Independent experiments could verify whether the results of our scaling procedure in a magnetic field can be interpreted in this way.

\subsection{Evaluation of the interaction parameter $\alpha_{ee}$}

The Coulomb interaction in a 2DES is modified owing to screening and correlation effects~\cite{stern}.
E-e scattering provides an important information about interactions via a scattering length (cross-section in 2D) $\lambda(\theta)$, where $\theta$ is a scattering angle~\cite{saraga}. The total scattering probability can be expressed through a total scattering length $\lambda_{tot}=\int_0^\pi{\lambda(\theta)d\theta}$.
In a classical approach of Refs.~\cite{NagaevAyvaz,NagaevKost,Nagaev_noise} this
is determined by a  dimensionless interaction parameter $\alpha_{ee}$: $\lambda_{tot}=\pi\alpha_{ee}/(4k_F)$. The measurement of the {\it e-e} scattering contribution to the PC resistance gives a direct access to the scattering length in the 2DES.

Eq.~(\ref{eq_Ayvaz}) permits determination of $\alpha_{ee}$ from the slope of the $T$-dependence at $B=0$, provided  the constriction half-width $a$ and the cutoff length $l_c$ are known. Moreover, analysis of the $B$-dependence of the {\it e-e} correction allows to exclude an uncertainty related to the unknown cutoff length $l_c$, as follows from eq.~(\ref{eq_kost}). However, application of this procedure to the present experiment is problematic owing to the discrepancies between theory and experiment discussed above. Still, we attempt to evaluate $\alpha_{ee}$ fitting the zero field data $\delta R(T)$ by a numerical calculation in spirit of Ref.~\cite{NagaevAyvaz}.

As discussed above, at temperatures used in our experiment the estimated mean-free path for {\it e-e} scattering is typically not large compared to the elastic mean-free path (and $l_c$), so that the electron beam decay should be taken into account. We employ fits with a numerical calculation, which accounts for the beam decay on the length-scale of the mean free path for the {\it e-e} scattering. Such fits are capable to describe the data within the whole $T$-range (see Appendix~\ref{numerics}) and serve as an order of magnitude estimate for $\alpha_{ee}$.


At a given $R_0$, the dependence $\delta R(T)/R_0$  from fig.~\ref{fig2}b is numerically fitted with three fit parameters $a,l_c$ and $\alpha_{ee}$.
The best fits are obtained for $l_c=5\mu m$ (for $R_0>0.64k\Omega$). We used a separate
value of $l_c=8.5\mu m$ when fitting the lowest PC-resistance data ($R_0=0.64k\Omega$), which best reproduces a downturn on the $T$-dependence (see fig.~\ref{fit_lee}). These values of the cutoff length are substantially smaller than the elastic mean-free path in our device ($l_0\approx20\mu$m). This discrepancy might arise from a crude model used and is not crucial for our purposes. Important for the evaluation of $\alpha_{ee}$ is the knowledge of the parameter $a$. For a classical PC $a$ simply equals the half-width of the orifice~\cite{NagaevAyvaz,NagaevKost}. However, this is not the case in present experiment for the following reason. In our T-shaped split-gates layout about a half of the electrons can reach the PC after a scattering off the 2DES boundary defined by the right-hand-side gate in the inset of fig.~\ref{fig1}. Assuming a specular boundary scattering, this results in a factor of 2 enhanced $a$ compared to the geometry of Ref.~\cite{NagaevAyvaz} at the same $R_0$. Therefore we perform a numerical fit with $a$ equal twice the physical half-width of the PC. We also perform an extra fit using effective width $a=2a_{eff}$, as suggested by the scaling of the experimental curves $\delta R(B)$ in a magnetic field.

\begin{figure}[t]
 \begin{center}
  \includegraphics[width=0.8\columnwidth]{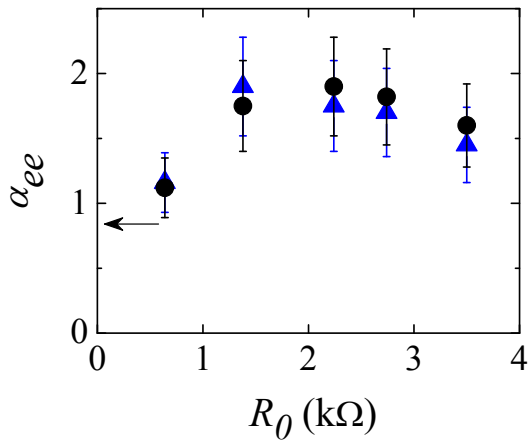}
   \end{center}
  \caption{Evaluation of $\alpha_{ee}$ from the zero field $T$-dependence. Circles and triangles correspond to fitting with $a$ determined by the physical width and the effective width, respectively, as explained in the text. The values of $a$ and $a_{eff}$ used for fitting are shown in fig.~\ref{fig7} as dots and crosses, respectively. Error bars ($\pm$20\%) mark the random uncertainty of the evaluation procedure. The arrow points at the value expected from the calculation~\cite{saraga} of the {\it e-e} scattering length (see text). }\label{fig8}
\end{figure}

The interaction constant obtained from such numerical fits is plotted in fig.~\ref{fig8} as a function of $R_0$ in the range $0.6k\Omega<R_0<3.5k\Omega$ (symbols). Circles and triangles correspond to fitting with $a$ determined by the physical width and the effective width, respectively. Error bars in the figure indicate an estimated random uncertainty of the evaluation procedure, which is mainly limited by that of the $R_0$ determination. For each data set $\alpha_{ee}$ is roughly independent of $R_0$ (exception is the case of $R_0=0.64k\Omega$), as expected for the parameter characterizing the strength of the {\it e-e} scattering in the 2DES. Based on fig.~\ref{fig8} we conclude that our experiment is consistent with an order of magnitude estimate~\cite{NagaevAyvaz} $\alpha_{ee}\sim1$. A direct calculation~\cite{saraga} of the 2D scattering length predicts comparable values
of the interaction parameter, which increases at decreasing the 2D electron density. We were able to check that at a density of $n_S=1.5\times10^{11}$cm$^{-2}$, still almost twice as high as in our 2DES, the calculations~\cite{saraga} predict $\alpha_{ee}\approx0.84$ (at $T=2$K). This value, shown by an arrow in fig.~\ref{fig8}, is roughly consistent with our evaluation. Unfortunately, a systematic uncertainty, associated with our fitting procedure and large number of fit parameters, by far exceeds the random uncertainty in fig.~\ref{fig8}. This doesn't permit a quantitative comparison beyond an order of magnitude estimate. The data of fig.~\ref{fig8} is rather meant to illustrate that in a wide range of $R_0$ the absolute value $\delta R(T)$ is roughly consistent with the predictions of the {\it e-e} scattering scenario.

\section{Summary\label{summary}}

In summary, we studied a $T$-dependent contribution $\delta R$ to the resistance of a ballistic PC in a 2DES of a high-quality GaAs/AlGaAs heterostructure at temperatures below 4.2K. In zero magnetic field, the resistance decreases by $\sim10-20$\% as the temperature is raised from 0.5K to 4.2K. The dependence is roughly linear below 2K and weakens at higher $T$. A $B$-driven suppression of $\delta R$ is found in perpendicular magnetic fields of a few 10mT and not too high $T$. These results give strong evidence for the influence of the {\it e-e} scattering on the PC conductance. The observations are similar in a wide range of $R_0$, even outside the classical PC regime, and can be qualitatively described with the {\it e-e} scattering scenario of Refs.~\cite{NagaevAyvaz,NagaevKost}. We argue that in the $B=0$ case the discrepancies between the experiment and theory can be well understood, and support by a numerical calculation. In magnetic field, the curves $\delta R(B)$ permit single-parameter scaling in a wide range of PC resistances $R_0$, which is an intrinsic property of the {\it e-e} scattering problem. Contrary to expectations, the dependence of the scaling parameter on $R_0$ is sufficiently weaker than that of the physical half-width of the orifice. This indicates that the value of the {\it e-e} scattering contribution in a lateral PC is determined by an independent parameter $a_{eff}$, which we call an effective half-width of the PC. In $B=0$ case, we perform a numerical calculation, which goes beyond theoretical assumptions~\cite{NagaevAyvaz} and is capable to quantitatively describe the experimental data in the whole temperature range. Using this calculation, the interaction constant $\alpha_{ee}$ of the 2DES is evaluated, which is in rough quantitative agreement with the calculations~\cite{saraga} of the {\it e-e} scattering length in 2DESs in GaAs.

\section{Acknowledgements}

We acknowledge discussions with I.L.~Aleiner, I.S.~Burmistrov, E.V.~Deviatov, V.T.~Dolgopolov, V.F.~Gantmakher, A.A.~Shashkin, D.V.~Shovkun.
We also wish to thank T.V. Krishtop and K.E. Nagaev for discussions and criticism.
Financial support from RAS, RFBR, the grant MK-3102.2011.2, the German Excellence Initiative
via the Nanosystems Initiative Munich (NIM) and LMUexcellent is gratefully acknowledged.
VP and LS wish to acknowledge financial support from Italian Ministry of Research through FIRBIDEAS Project No. RBID08B3FM.

\appendix

\section{Evaluation of the electron density inside the PC\label{depopulation}}

The resistance  $R_0$ of a quasiclassical PC and its physical half-width $a$ (at a Fermi level) are directly related as
$R_0\propto C_1/(ak_F^{PC})$, where the Fermi wave-vector is given by $k_F^{PC}=(2\pi n^{PC})^{1/2}$ and $n^{PC}$ is the (2D) electron density inside the PC. The numerical coefficient depends on the model of the confinement potential and equals $C_1=\pi$ (4), respectively, for a hard-wall (parabolic) confinement.
Separate determination of $a$ and $n^{PC}$ is possible by analyzing the depopulation of the magneto-electric subbands in a perpendicular magnetic field.

The depopulation effect gives rise to a positive contribution to the magnetoresistance of the  quasiclassical PC, which
to the lowest order in $B$ is given by~\cite{vanHouten_beenakker}:
\begin{equation}\label{depop}
    \frac{\delta R_{depop}}{R_0}=C_2\left(\frac{a}{R_C^{PC}}\right)^2\propto (C_1)^2C_2\frac{B^2}{(n^{PC})^{2}},
\end{equation}
where $R_C^{PC}=k_F^{PC}\hbar c(eB)^{-1}$ is the cyclotron radius corresponding to the electron density inside the PC. The confinement model dependent coefficient equals $C_2=1/6$ (1/2), respectively, for a hard-wall (parabolic) confinement. Here we also used $a\propto C_1/k_F^{PC}$, since $R_0$ is fixed. The evaluated density depends on the details of the PC confinement via a factor $C_1(C_2)^{1/2}$ equal to $\approx1.28$ (2.83) for a hard-wall (parabolic) model confinement potential. In fact, a gradual transition from a hard-wall-like confinement in a wider constriction to a parabolic-like confinement in a narrower constriction is typically found~\cite{Wharam_confinement} in agreement with numerical simulations (see~\cite{vanHouten_beenakker} for references). Here, we evaluate the electron density $n^{PC}$ assuming a hard-wall confinement with the help of eq.~(2.62a) of Ref.~\cite{vanHouten_beenakker}. In this way a lower boundary for $n^{PC}$ is obtained and the strongest possible gate-voltage dependence of the electron density inside the PC.

\begin{figure}[t]
 \begin{center}
  \includegraphics[width=0.8\columnwidth]{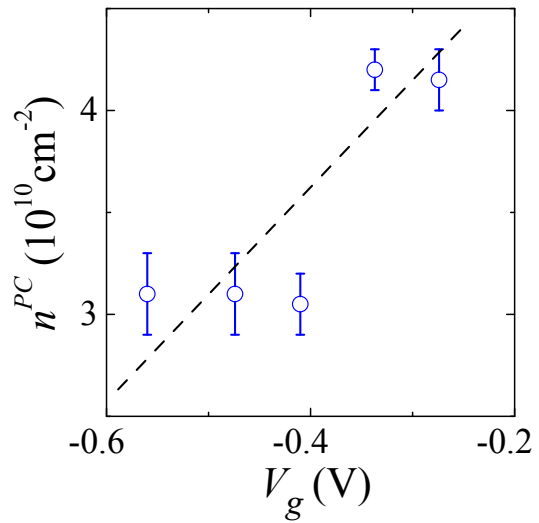}
   \end{center}
  \caption{Evaluation of the electron density inside the PC from subband depopulation effect, assuming a hard-wall confinement. Experimental points for a set of gate-voltages are shown by symbols. These symbols indicate values of the density averaged over several experiments at the same $V_g$.
  Error bars represent a random uncertainty, which comes from both the accuracy of the fits and the reproducibility of the magnetoresistance traces $R(B)$. The dashed line is the best linear fit  $\delta n^{PC}\propto \delta V_g$. }\label{fig10}
\end{figure}

The result is shown in fig.~\ref{fig10} for gate voltages in the range corresponding to $0.64k\Omega\leq R_0\leq3.5k\Omega$ (symbols). Here, the lowest $T$ experimental data were used, where the contribution of the {\it e-e} scattering to the magnetoresistance is not important (see section~\ref{Bfield}).
The evaluated density $n^{PC}$ is reduced by at least a factor of 2 compared to that in the bulk 2DES and further decreases towards the lower $V_g$.
Note that assuming a parabolic confinement would result in factor of 2.2 higher density, that is pretty close to the bulk 2DES value ($n_S\approx0.83\cdot10^{11} cm^{-2}$). Most probably, the actual density is somewhere in between the two limiting cases.

Strictly speaking, the data scatter in fig.~\ref{fig10} precludes an accurate evaluation of the gate-voltage dependence of $n^{PC}$. As a crude estimate, we take the simplest capacitive approximation $\delta n^{PC}\propto \delta V_g$ to describe the data, which is reasonable for a relatively deep 2DES used. The dashed line is the best linear fit extrapolating to $n^{PC}=0$ near the pinch-off point of the PC at $V_g\approx-1.1$V. It is this fit, that was used for the evaluation of the physical half-width $a$ in fig.~\ref{fig7}. Note that a $\approx16\%$ uncertainty in $a$, associated with the confinement model choice ($a\propto C_1^{1/2}C_2^{-1/4}$), is not crucial for us (see fig.~\ref{fig8}). At the same time, an account of a gradual  change from a hard-wall-like to a parabolic-like confinement at decreasing $V_g$ could only cause a functional dependence of $a(R_0)$ in fig.~\ref{fig7} to become farther from $a_{eff}(R_0)$.

\section{Numerical calculation\label{numerics}}

We perform a numerical calculation of the {\it e-e} scattering contribution in a 2DES for $B=0$ in the spirit of Ref.~\cite{NagaevAyvaz}.
This task involves a 6-dimensional integration:

\begin{eqnarray}
\nonumber  \frac{\delta G}{G_0}=\frac{\alpha_{ee}k_Fa}{8\pi^2}\int_0^{\pi/2}\cos\phi_p d\phi_p\int_1^{l_c/a}dr  \\
\label{num_int}
\\
\nonumber
\int d\phi_{k}\int d\theta\int\int\frac{f_{p'}f_{k'}+f_p-f_pf_{p'}-f_pf_{k'}}{T\cosh^2\frac{\varepsilon_k-1}{T}}d\varepsilon_pd\varepsilon_k
\end{eqnarray}

Here $p,\varepsilon_p$ ($k,\varepsilon_k$) is the momentum and energy (in units of $E_F$) of the electron incident to (injected through) the PC orifice, respectively. After scattering the electrons acquire the momenta $p'$ and $k'$. The angle of $p$ in respect to the direction normal to the orifice is denoted as $\phi_p$; $\phi_k$ is the angle between the direction towards the center of the orifice and momentum $k$; $\theta$ is the scattering angle in the relative momentum space. The distance $r$ from the PC is measured in units of PC half-width $a$ (region $r<a$ is omitted) and the temperature $T$ -- in units of $E_F$. The equilibrium distribution functions for the momenta $p,p',k'$ are denoted by $f_p,f_{p'},f_{k'}$. In this expression we accounted for the energy and momentum conservation laws and substituted the distribution $f_k$  by its nonequilibrium part linearized in respect to bias voltage across the PC. Integration over $\phi_k$ is restricted to the angular dimension of the electron beam injected through the PC $\phi_k\in(-\phi_{max},\phi_{max})$, where $\phi_{max}=\arctan(\cos\phi_p/r)$ . Integration over $\theta$ is limited to the scattering angles such that electrons $k'$ and $p'$ don't reach the orifice. At a given $T$ we restrict the integration over electrons' energies to the interval $E_F\pm5T$.

\begin{figure}[t]
 \begin{center}
  \includegraphics[width=0.8\columnwidth]{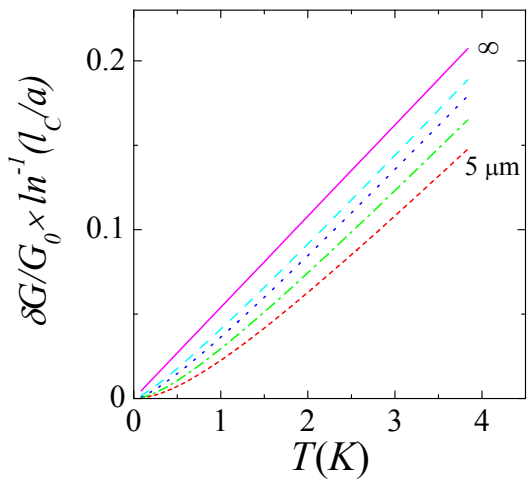}
   \end{center}
  \caption{Results of a numerical calculation of the {\it e-e} scattering contribution in zero magnetic field. The data correspond to $\alpha_{ee}=1, a=400$nm and
a set of $l_c=\infty,100,30,10,5\mu$m from top to bottom curve. The ordinate axis is chosen to illustrate the dependence on $l_c$, see text.}\label{fig_num}
\end{figure}

In fig.~\ref{fig_num} we plot the results of the numerical calculation for $\alpha_{ee}=1, a=400$nm and a set of $l_c$ values between 5$\mu$m and 100$\mu$m. We find that at high $T$ the slope of the $T$-dependence is approximately proportional to $\ln(l_c/a)$, in agreement with the analytic result of eq.~(\ref{eq_Ayvaz}). To illustrate this, for each $l_c$ the calculated relative contribution $\delta G/G_0$ was divided by $\ln(l_c/a)$ in fig.~\ref{fig_num} to give approximately the same slope at high $T$. In the limiting case $l_c\rightarrow\infty$ we recover the result of eq.~(\ref{eq_Ayvaz}), although the numerical coefficient obtained is a factor of $\approx$1.7 larger. The origin of this discrepancy is unclear and we compensate for this when fitting the experimental data.

\begin{figure}[t]
 \begin{center}
  \includegraphics[width=0.8\columnwidth]{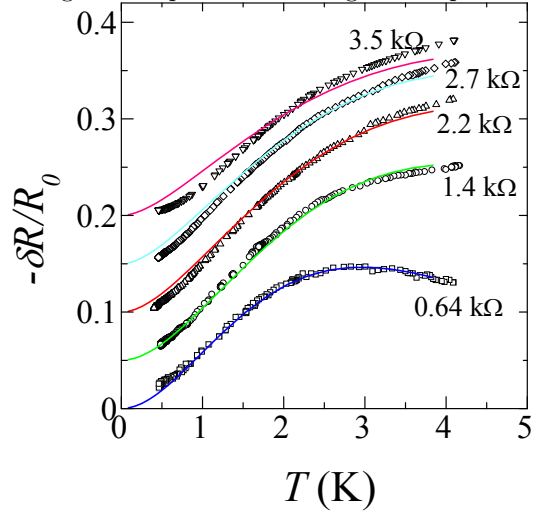}
   \end{center}
  \caption{Normalized experimental $T$-dependence of the PC-resistance (symbols) together with numerical fits accounting for the beam decay effect (lines),
 shifted in steps of 0.05 along the ordinate axis for clarity. These fits are obtained with the cut-off length equal to $l_c=8.5\mu$m for $R_0=0.64k\Omega$ and $l_c=5\mu$m for the four other curves. The parameter $a$ equals twice the physical half-width of the PC orifice from fig.~\ref{fig7} (dots). The {\it e-e} interaction parameter values are given in fig.~\ref{fig8} (circles).}\label{fit_lee}
\end{figure}

 With increasing $T$ a mean-free path $l_{ee}$ for the {\it e-e} scattering decreases and eventually becomes comparable to or smaller than the cut-off length scale $l_c\sim l_0$. This means that an electron beam injected through the PC decays at a typical distance of $l_{ee}$ from
the PC orifice. In order to numerically estimate the {\it e-e} scattering contribution to the PC-resistance in this regime we weight the integral over the dimensionless distance $r$ in~(\ref{num_int}) by an exponential factor $\exp(-ar/l_{ee})$, where the $T$-dependent $l_{ee}$ is taken from the quasiparticle life-time calculations~\cite{giuliani}. Most probably, this approach can not be justified by a solution of the kinetic equation and should be treated as a first crude step on the way to a self-consistent calculation. Nevertheless, such an account of the beam decay allows to describe the experimental $T$-dependence in the whole range of temperatures. In fig.~\ref{fit_lee} we plot the best numerical fits (lines) to the experimental data (symbols) for the 5 lowest values of $R_0$ used. The data ordinate are shifted in steps of 0.05 for successive $R_0$. The saturation of the $T$-dependence in the range $T>2$K is perfectly reproduced by the fits, which strongly supports our interpretation in terms of the {\it e-e} scattering. The overall quality of the fits is pretty good for $R_0\leq2.7k\Omega$. At higher $R_0$ the agreement is worse, presumably thanks to a residual thermal smearing of the conductance plateaus (see section~\ref{zerofield}).

As follows from fig.~\ref{fig_num} and fig.~\ref{fit_lee}, a cut-off length on the order of $l_c\sim10\mu$m is too small to clearly observe the proportionality~\cite{NagaevAyvaz} $\delta G_{ee}\propto T$. Improving the sample quality one could achieve longer elastic mean-free path (and $l_c$), which favors linearity of the $T$-dependence at temperatures such that $T/E_F\gg a/l_c$ (see fig.~\ref{fig_num} and discussion in section~\ref{disc_zeroB}). On the other hand, the beam decay effect could not be neglected unless $l_c\ll l_{ee}$, which is harder to satisfy at higher $l_c$. The trade-off is to increase $l_c$ and measure at lower temperatures in order to satisfy simultaneously $a(T/E_F)^{-1}\ll l_c\ll l_{ee}\propto(T/E_F)^{-2}/\ln(E_F/T)$. This conditions are possible, yet challenging to meet, and is simpler to do in higher density (large $E_F$) samples. For a GaAs, we expect that a $T$-dependence close to $\delta G_{ee}\propto T$ could be observed at $0.1{\rm K}<T<1$K in samples with a 2DES mobility exceeding $10^7$cm$^{2}/$Vs.


\begin{thebibliography}{50}
\bibitem{roundingfig3a} A sharp $B=0$ kink of the dashed line in the inset of fig.~\ref{fig3}a is a property of an idealized 2DES. A finite
elastic mean-free path of $l_0\approx20\mu$m in our device is believed to result in a rounding of the kink in magnetic fields of a few mT, where the cyclotron radius exceeds $l_0$.
\bibitem{WL} This behavior in magnetic fields is in stark contrast with that of a well-known
weak-localization (WL) correction in diffusive 2D and quasi-1D systems. In those systems, the low-$B$ magnetoresistance
is (i) negative and (ii) strengthens at decreasing $T$ (see, e.g., Ref.~\cite{Choi_WL} and Ref.~\cite{vanHouten_beenakker} for a review), which is opposite to our observations. Obviously, the WL physics is not important in clean ballistic PCs we study here.
\end{thebibliography}
\end{document}


\title{Supplementary material}

\author{M.~Yu.~Melnikov}
\affiliation{Institute of Solid State Physics, Russian Academy of
Sciences, 142432 Chernogolovka, Russian Federation}
\author{J.P.~Kotthaus}
\affiliation{Center for NanoScience and Fakult\"{a}t f\"{u}r Physik,
Ludwig-Maximilians-Universit$\ddot{\text{a}}$t,
Geschwister-Scholl-Platz 1, D-80539 M$\ddot{\text{u}}$nchen,
Germany}
\author{V. Pellegrini}
\affiliation{NEST, Istituto Nanoscienze-CNR and Scuola Normale Superiore, Piazza San Silvestro 12, I-56127 Pisa, Italy}
\author{L. Sorba}
\affiliation{NEST, Istituto Nanoscienze-CNR and Scuola Normale Superiore, Piazza San Silvestro 12, I-56127 Pisa, Italy}
\author{G.~Biasiol}
\affiliation{CNR-IOM, Laboratorio TASC, Area Science Park, I-34149 Trieste, Italy }
\author{V.S.~Khrapai}
\affiliation{Institute of Solid State Physics, Russian Academy of
Sciences, 142432 Chernogolovka, Russian Federation}

\maketitle

\section{Low temperature conductance quantization}

\begin{figure}[t]
 \begin{center}
  \includegraphics[width=0.8\columnwidth]{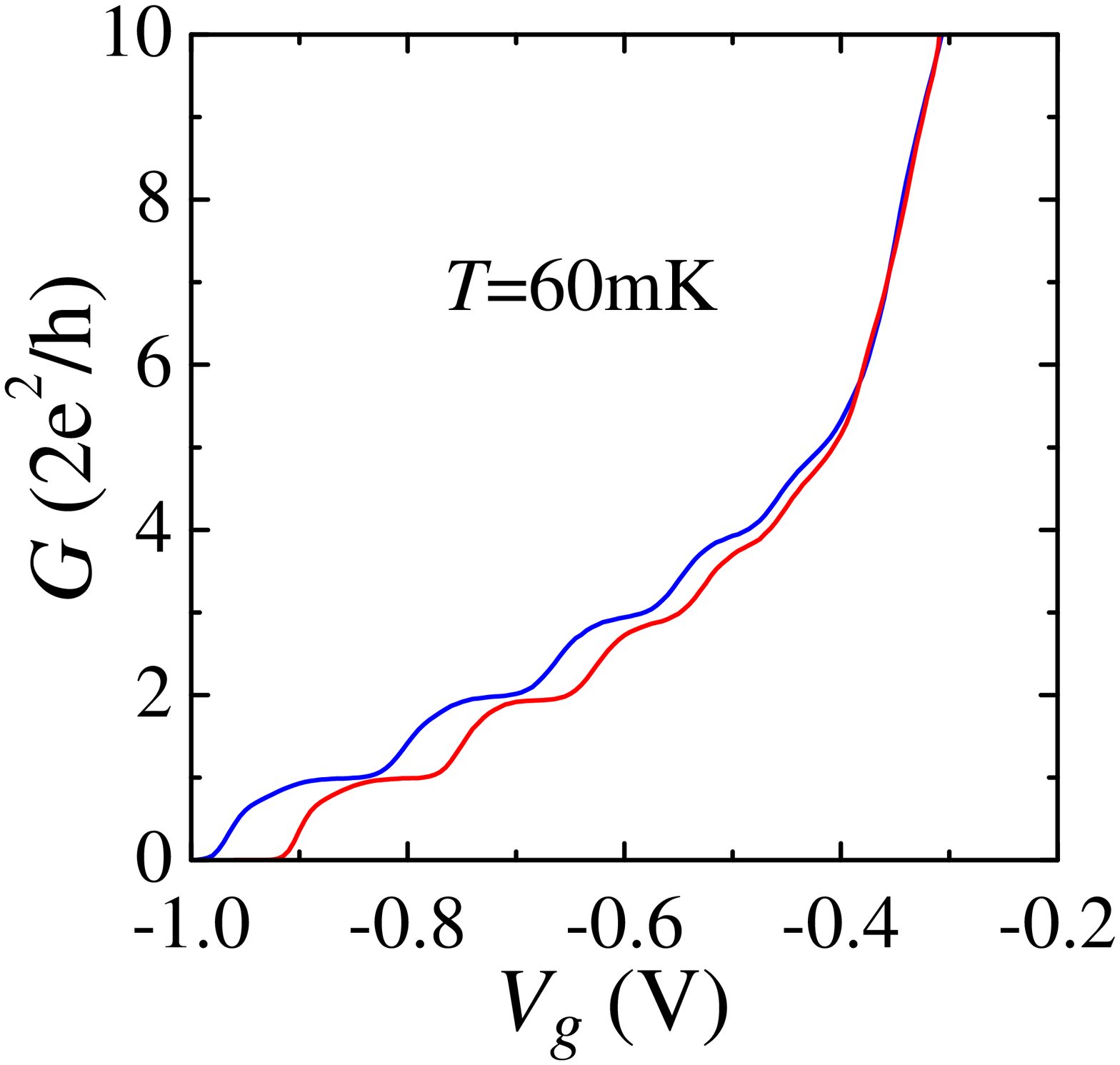}
   \end{center}
  \caption{Gate-voltage dependence of a linear-response PC-conductance in one of the samples studied in the main paper (two different constrictions).
  The data are taken in a dilution $^3$He/$^4$He refrigerator with a 60mK base temperature.}\label{fig_sup_1}
\end{figure}

We study the {\it e-e} scattering contribution to the point-contact (PC) resistance at
temperatures above $T\approx0.5$K. As shown in fig.~1 of the main paper,
the resistance quantization is rather poor, which we interpret as a result
thermal smearing of the Fermi distribution function~\cite{Bagwell}. In order to demonstrate this,
we cool one of the samples down to $T\approx60$mK and observe clearly quantized conductance plateaus, see fig.~\ref{fig_sup_1}.
Such a quality of quantized PC conductance is routinely observed in our devices at low-$T$ and rules out disorder as origin of the
poor quantization at higher temperatures.

\section{$R_0$ at elevated temperatures}

\begin{figure}[t]
 \begin{center}
  \includegraphics[width=0.8\columnwidth]{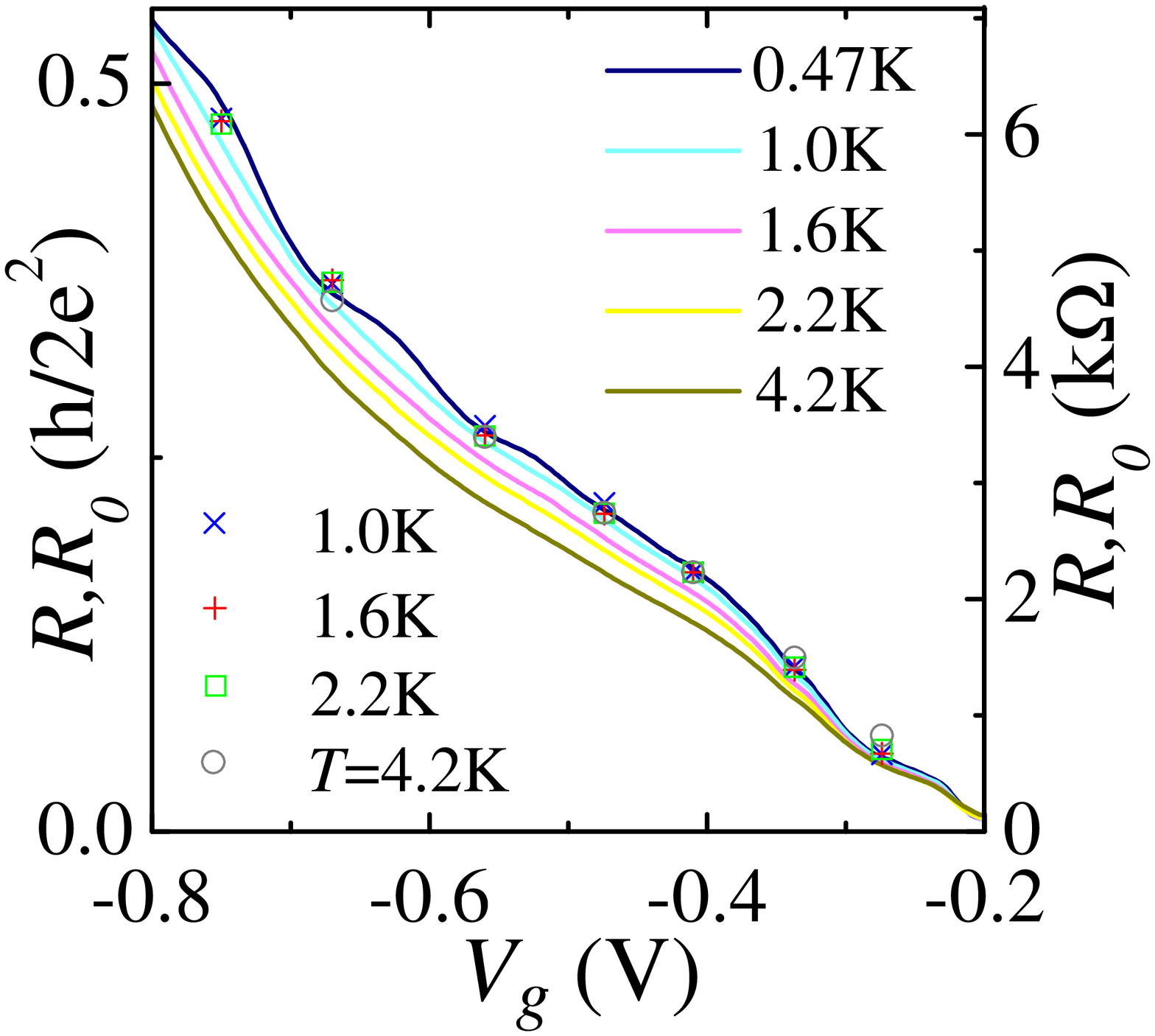}
   \end{center}
  \caption{Gate-voltage dependence of the PC-resistance for a set of temperatures used in experiment (lines, see legend). Symbols mark the values of $R_0$ used for scaling of the magnetoresistance data in fig.~6b of the main paper at temperatures $T\geq1$K (see legend). }\label{fig_supl_2}
\end{figure}

As discussed in the main paper, a direct extrapolation of the $T$-dependence $R(T)$ towards a zero temperature limit ($R_0$) is problematic.
One way to determine $R_0$ is from a magnetoresistance data in small magnetic fields at the lowest $T$, where the  {\it e-e} scattering contribution is almost absent. However, this procedure is not applicable at $T\geq1$K, where magnetoresistance is not of purely single-particle origin. Unfortunately, resistance drifts on the order of $\pm$2\% prevent using the lowest-$T$ value of $R_0$ at a given gate-voltage. Instead, when scaling the magnetoresistance data, we allow  a variation of $R_0$ by a few percent at a given $T$ and $V_g$. In fig.~\ref{fig_supl_2}, the $R_0$ values used for scaling in fig.~6b of the main paper (symbols) are plotted along with the measured gate-voltage dependencies of the PC-resistance for a set of temperatures studied (lines). The $R_0$ data points exhibit scatter at different $T$ and fall close to the lowest temperature trace $R(V_g)$, except for some points at $T=4.2$K.

